\documentclass[prd,twoside]{revtex4}

\def\be{\begin{equation}}
\def\ee{\end{equation}}
\def\la{\label}
\def\bea{\begin{eqnarray}}
\def\eea{\end{eqnarray}}
\def\non{\nonumber}
\def\ci{\cite}
\def\la{\label}
\def\bib{\bibitem}

\def\lm{\lambda}

\def\le{\left}
\def\ri{\right}

\def\al{\alpha}
\def\w{\varpi}

\def\D{\Delta}
\def\rp{\rho_\phi}
\def\rpo{\rho_{\phi o}}
\def\wp{w_\phi}

\def\wpo{w_{\phi o}}

\def\s8{\sigma_8}

\def\fr{\frac}
\def\pp{\partial}
\def\pu{\pp_\mu}
\def\pU{\pp^\mu}

\def\half{\fr{1}{2}}

\usepackage[latin1]{inputenc}
\usepackage{graphicx}
\usepackage{psfrag}

\begin{document}

\title{Causality and Sound Speed for General Scalar Field Models,
 including $w < -1$, tachyonic,  phantom,  k-essence and curvature corrections }

\author{Axel de la Macorra}
\affiliation{Instituto de F\'{\i}sica,
UNAM, Apdo. Postal 20-364, 01000 M\'exico D.F., M\'exico}
\email{macorra@fisica.unam.mx}
\author{Héctor Vucetich}
\altaffiliation[On leave of absence from]{ Facultad de Ciencias
  Astronómicas y Geofísicas, Universidad Nacional de La Plata,
  Argentina.}
\email{vucetich@fisica.unam.mx}
\affiliation{Instituto de F\'{\i}sica, UNAM, Apdo. Postal 20-364,
01000 M\'exico D.F., M\'exico}

\begin{abstract}
The result from the SN1a projects suggest
that the dark energy can be represented by
a fluid with $w<-1$. However, it is commonly argued that a fluid
with $|w|>1$ contradicts causality.

Here, we will show that a fluid with $|w|>1$ does not contradict
causality if $w$ is not constant. Scalar field are the most
promising candidates for describing the dark energy and they do
not have a constant equation of state parameter $w$. Scalar
potentials may lead to regions where $w $ is larger or smaller
than one and even regions where the group velocity $d p/d\rho$
diverges.

We study the evolution of scalar field perturbations and we show
that the "sound speed" is always smaller than the speed of light
independently of the value of $w=p/\rho$ or $d p/d\rho$. In general,
it is neither the
phase velocity nor the group velocity that
gives the "sound speed".
 In the analysis we include the special cases of $|w|>1$, tachyonic,
phantom,   k-essence scalar fields and curvature corrections. Our  results   show that
 scalar fields do not contradict causality as long
as there is dispersion.

\end{abstract}


\maketitle

The existence of a dark  energy,  an energy density $\rho$ with
negative pressure  $p$ and equation of state $w=p/\rho$,  has been
determined by the SN1a \ci{SN1a} and the CMBR observations
\ci{CMB}. These observations show that we are living in a flat
universe with a matter contribution today $\Omega_m\simeq 0.3$ and
a dark energy $\Omega_{de}\simeq 0.7$ with $w_{de}\leq -2/3$. The
cosmological results do not rule out an energy density with $w<-1$
\ci{sn}.

Perhaps the most economic solution to the dark energy
is that of a scalar field. Not only because they are widely
predicted by particle physics but also because their
dynamics may lead in a natural way to an accelerated
universe. These scalar fields are called quintessence
and they are homogeneous in space and have only gravitational
interaction with all other fields.  Other interesting possibilities
 have been consider such as
phantom fields \ci{cald}, k-essence \ci{kessence} or tachyonic
scalar fields and curvature corrections \ci{cur-Vuc}-\ci{cur2}.

The possibility of having a fluid with $w<-1$ is problematic.
There are objections against such a fluid from vacuum instability,
causality and sound speed among other considerations. For a fluid
with a constant equation of state parameter $w$ it cannot exceed
the value of $|w|>1$. One of the many  problems appears as the
propagation of sound, since for $w$ constant it gives the sound
speed and clearly $w$ must be smaller than the speed of light
$c=1$ \ci{carter}. Another related problem is the vacuum
instability for fluids with a constant $w<-1$ \ci{insta}. Here, we
are interested in studying the problem of causality and sound
speed for fluids with $|w|>1$ or $|dp/d\rho|>1$. As we will see in
the next section a scalar field with $|w|>1$ and/or $|dp/d\rho|>1$
can be easily obtained if the field has negative potential
\ci{ax.neg} (e.g. a tachyon, $m^2<0$) or for a phantom field with
negative kinetic energy and positive potential \ci{cald}.
Concerning the evolution of the perturbations of the scalar field
we will show that in the absence of  gravity a tachyonic
field is equivalent to a phantom field.

We will
show that the information speed for a scalar field does not
contradict special relativity (see \ci{mcin}) even in the cases
where $|w|>1$ and $|dp/d\rho|>1$. However, in these cases
the amplitude of the perturbations grows rapidly and therefore
the first order approximation used to study the sound speed
fails. This signals that the space (universe) is no longer
homogenous and non-perturbative  mathematical methods must be
used to analyze the behavior of the universe. Still, we want
to emphasis that scalar fields with
$|w|>1$ and $|dp/d\rho|>1$, tachyons, phantom or k-essence models
do not violate causality and the sound speed is smaller or equal to
the speed of light.

\section{Motivation}

It is usually stated that the "speed of sound" for a fluid is
given either by $\wp\equiv p/\rho$ if it is constant or by $dp_o/d
\rho_o$ if $\wp$ is time dependent, where $p_o$ is the
pressure and $\rho_o$ the energy density.  In the case that $w\equiv p/\rho$ is
constant then the two quantities $dp/d\rho=w$ coincide.  The
requirement that the propagation of information should not exceed
the speed of light implies that $|\wp|<1$ and/or $ |dp_o/d\rho_o|<1$.
However, as we will show, neither of these two conditions are
necessarily satisfied for a scalar field.

Considering an homogenous  scalar field,
it is well known  that its evolution
leads naturally to a non-constant $\wp$,
so $\wp$ should not be interpreted as the
"speed of sound". The equation
of state parameter for a scalar field is
\be\la{wp}
\wpo\equiv\fr{p_{\phi o}}{\rpo}=
\fr{\lm\half \dot\phi_o^2-V(\phi_o)}{\lm\half \dot\phi_o^2+V(\phi_o)}
\ee
where the $o$ denotes that we are considering
an homogenous scalar with $\phi_o(t)$ and $\dot\phi\equiv d\phi/d t$
and we have included the parameter $\lm=\pm 1$ to take
into account a canonical scalar field $\lm=1$ and a phantom field $\lm=-1$.
It is easy to see from the above equation that
for any scalar field with positive
kinetic term $E_k=\dot\phi^2/2\geq 0$
that  the magnitude and sign $\wp$
depends on the value of the potential $V(\phi)$
and eq.(\ref{wp}) gives
\bea
|\wpo| \leq 1  & \Leftrightarrow  & V(\phi_o)\geq 0  \\
 \wpo > 1       & \Leftrightarrow & E_k>-V>0 \non \\
\wpo < -1 &\Leftrightarrow         & E_k+V<0.
\la{const.wp}
\eea
As we see from eqs.(\ref{const.wp}) we can  obtain $\wpo<-1$ for
a scalar field with a negative potential (e.g. for a tachyon
field with $V=m^2\phi_o^2,\;m^2<0$). We can
also have $\wpo<-1$ for a phantom field with
negative kinetic energy ($E_k<0$) with positive potential ($V(\phi_o)>0$).
 It was shown in \ci{ax.neg} that this kind of behavior  is reached
naturally for scalar field with a negative minimum in the presence
of a barotropic fluid  that can be, for example, matter or
radiation. Therefore, we do not need exotic
fluids to have regions with $|\wp|>1$.

Now, let us analyze the quantity $dp_\phi/d\rho_\phi$ for a scalar
field with arbitrary potential. The homogenous part of the scalar
field depends only on time and using eq.(\ref{wp}) we have the
adiabatic quantity
\be\la{cs}
\fr{d p_{\phi}}{d \rho_\phi}=\fr{\dot p_\phi}{\dot\rp} = 1-\fr{2 V'}{\lm\ddot\phi+V'}=
1+\fr{2}{3}\fr{V'}{\lm H\dot\phi}
\ee
where we have used
\be\la{dp}
\dot p_\phi=\dot\phi(\lm\ddot\phi-V')\hspace{2cm}
\dot\rp=\dot\phi(\lm\ddot\phi+V')
\ee
and the equation of motion of the
scalar field
\be
\lm(\ddot{\phi}+3H\dot\phi)+V'=0
\ee
for the second
equality of eq.(\ref{cs})  with
 $V'\equiv dV/d\phi$. Notice that in general eq.(\ref{cs}) has not
only regions where $|dp_\phi/d\rho_\phi|$ is larger than one but
regions where it diverges. This will happen at the turning points
of the field oscillations around
 the minimum of the potential
when $\dot\phi=0$, if the point is reached at a non-extremum of
the potential $V$, (i.e. $\lm\ddot\phi=-V'\neq 0$), e.g. for any potential with
a minimum at finite $\phi$ as for a massive scalar field with
$V=m^2\phi^2$.

Clearly, we see that under normal conditions
both (homogenous) quantities $\wp, dp_\phi/d\rho_\phi$ can be larger
than the one (the speed of light).
Does this imply that the sound speed of scalar
fields travel faster then the speed of light?
The answer is no, as we will show later,
and the reason is that neither   $\wp$ nor $ dp_\phi/d\rho_\phi$
gives the correct interpretation
of the sound speed.

\section{Summary of Different Velocities}

In order to clarify the notion of causality for fluid
or a scalar field it will be useful to define
the different velocities involved in determining
the transportation of a signal in a medium.

For simplicity in presentation purposes we
will consider only one space dimension.
The wave equation for a fluid is given
by
\be\la{onda0}
\ddot\rho-\fr{v^2}{c^2}\;\;\pp_x^2 \rho + d\; \dot\rho  +b \rho =0
\ee
where $c$ is the speed of light and the parameters
$v, b, d$ may depend, in general, on the coordinates $t,x$
and they define the type of medium the fluid is in.
If we take a Fourier transformation
\be\la{rk}
\rho(t,x)=\int_{-\infty}^{\infty} dk \rho_k(t) e^{ik x}
\ee
eq.(\ref{onda0}) takes the form
\be\la{onda}
\ddot\rho_k  + d\; \dot\rho_k  +(\fr{v^2}{c^2}k^2+b)\; \rho_k =0.
\ee

The parameter $v$ defines the phase velocity while the friction term
is given by $d$.
The simplest wave equation is given by  $d=b=0$ and
$v$  constant. In this case  all monochromatic wave functions
will have the same speed and $v$  can be interpreted as the sound
speed. For $d=0$ and $(v^2k^2/c^2+b)/k^2\;$ is
$k$-dependent, the medium is said to be dispersive and
the monochromatic wave functions will travel
with different phase velocities. In this case
it is the group velocity that determines the
speed of sound. If we allow $d$ to be different
than zero then we have a dissipative medium
and the wave amplitude will suffer
a damping or growth depending on the
sign of $d$ and neither the phase nor the group
velocity would give the "sound speed".

For $d=b=0$ and $v=cte$ the solution to eq.(\ref{onda}) involves
a sum of monochromatic waves
\be\la{r2}
\rho(t,x)=\int_{-\infty}^\infty \le(A_k\;e^{-i(k x-\w t)}+B_k\;e^{i(k
  x-\w t)}\ri)dk
\ee
with $\w,k$ constants and the amplitudes $A_k,B_k$ also constant.
From eq.(\ref{onda}), with $b=d=0$, the frequency $\w$ (it should
not be confused with $w$ the equation of state
parameter defined in the previous  section) is a function of k
\be\la{wk}
\w^2=\fr{v^2}{c^2}k^2.
\ee
The wave number is given by $k=2\pi/\lm$ and gives
the inverse of the wave length.
From now on we will set the speed of light back to one ($c=1$).
The speed for a monochromatic wave, e.g. the speed of the maximum
of the wave, is given by the constrain $kx-\w t=cte$ from which we
have the phase velocity
\be\la{vp}
v_p\equiv\fr{dx}{dt}=\fr{\w}{k}=v
\ee
if $k, \w$ are independent of $x,t$.
 A  velocity smaller than $c$ requires
$v_p=\w/k<1$. Notice that in this case $v_p$
is the same for all monochromatic wave functions
since $w/k$ is $k$-independent.

If we are in a dispersive medium $b\neq 0$ and
the wave frequency depends on the value of $k$.
In this case one needs to consider the evolution
of wave packets since their shape will be time dependent.
A wave packet occupies a limited region in space. It is common
to assume that the monochromatic frequencies that form
the packet are
concentrated around a central value $\w$
with corresponding wave number $k$.
For a wave function with a slightly different wave number
$k'=k+\D k$ the corresponding frequency is $\w(k+\D k)$
and for $\D k $ small we can write $\w(k+\D k)\simeq \w(k)+(d\w/dk) \D
k$.
The solution to the wave equation for a wave packet is of the form
\be
\rho(t,x)=A \;e^{i(kx-\w t)}f ((\D k x-t\D \w))
\ee
and the amplitude of the wave is modulated by the function
$f$. The group velocity is defined by requiring the argument
of  $f$ to be constant,
i.e. by the solution of $\D k x-t\D \w=cte$,  giving the group
velocity
\be\la{vg}
v_g=\fr{\D \w}{\D k}=\fr{d\w}{dk}
\ee
if $d\w/dk$ is $t,x$ independent. The group velocity $v_g$
is valid for any dependence of $\w(k)$ but only as long
as $\D k$ is small.
If the medium is non dispersive
$\w$ is constant then the group velocity and the phase
velocity coincide $v_g=v_p=\w/k$.

In general the form of the packet will
not remain constant and it will be smoothed out during its propagation
and the condition of having a small $\D k$ will no be
maintained.

Lastly let as consider a dissipative medium in the simple
case of constant $b,d,v$. In this case the solution
to eq.(\ref{onda}), with the ansatz $\rho_k=Ae^{i\w t}+Be^{-i\w t}$
give the equation,
\be
-\w^2+id\w+(\fr{v^2}{c^2}k^2+b)=0
\ee
with
\be
\w=\fr{id\pm\sqrt{4(\fr{v^2}{c^2}k^2+b)-d^2}}{2}
\ee
and we see that $\w$ takes imaginary values if $d\neq 0$
or $4(k^2v^2/c^2+b)-d^2<0$. This will give
damping or a  growing wave function solution.
If $4(k^2v^2/c^2+b)-d^2>0$ the damping term will
be the same for all monochromatic modes.

\begin{figure}[tbp]
  \begin{center}
    \includegraphics{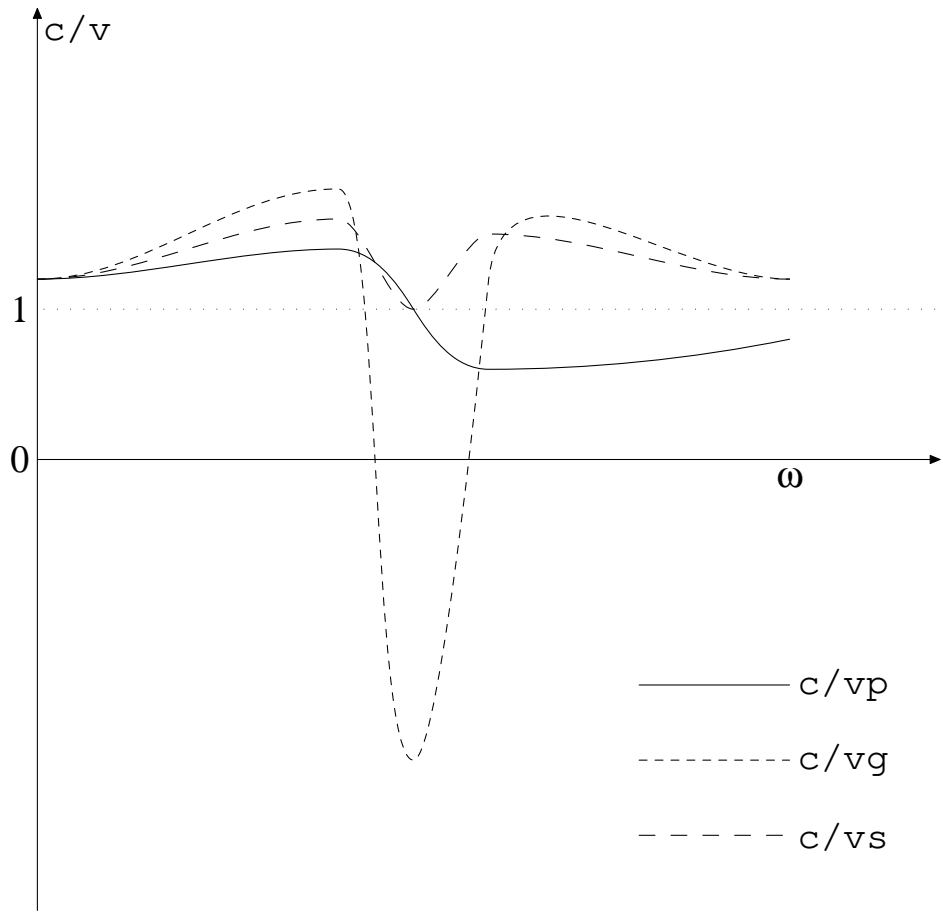}
  \end{center}
  \caption{Propagation speeds near an absorption band. Although the
  phase velocity becomes larger than light speed and the group
  velocity becomes infinite and even negative, the signal velocity is
  always smaller than $c$.}
  \label{fig:SomBrill}
\end{figure}

The short explanation we have given above also shows conditions for the
validity of the approximation: the packet must be quasi-monochromatic
and the frequency a slowly varying real function of $k$. If either of
these approximations breaks down, the group velocity loses its
physical meaning. In the general case the "speed of sound" is neither
the group nor the phase velocity and one must use the more complex and
subtle notion of signal velocity \cite{Sommerfeld,Brillouin}.
 The signal velocity is defined as the velocity with which a given
standard \emph{amplitude} of the wave packet moves, for instance half
that of the maximum amplitude. The standard Sommerfeld-Brillouin
definition amounts to this. The condition of \emph{retarded} wave
guarantees then that the signal velocity will be always smaller than
$c$, even if the wave packet deforms very much and the group velocity
becomes $v_g > c$. That this may happen is shown in the classical
Sommerfeld-Brillouin diagram \cite{Stratton} for the three velocities
in the neighborhood of an absorption line (Fig. \ref{fig:SomBrill}).
In the case of a scalar field the group velocity diverges
when $\dot\phi=0$ with $V'\neq 0$. This will happen in most
cases and even for a potential of the form $V=m^2\phi^2$ around the
turning points of the scalar field.

\section{Perfect Fluid and Scalar Field }

\subsection{Perfect Fluid}

A perfect fluid has an energy momentum tensor
$T^{\mu\nu}\equiv (\rho+p)u^\mu u^\nu-pg^{\mu\nu}$
where $\rho$ is the energy density and $p$ the pressure
of the fluid, $g^{\mu\nu}$ the metric tensor while $u^{\mu}$
the 4-velocity. In the rest frame $u^o=1, u^i=0$ and
$T^{\mu\nu}$ gives the usual energy momentum tensor
$T^\mu_{\;\;\nu}=diag[\rho,-p,-p,-p]$. For an arbitrary
velocity $v$ the four velocity is $u^\mu=u^o(1,v^i)$
with $u^0=1/\sqrt{1-v^2}$. The equation of motion are given
by the conservation
equation $ \pp_\mu T^{\mu\nu}=0$
and to solved them one needs to add
an equation of state $p=w\rho$. Matter
and radiation have a constant $w$ equal
to zero, 1/3 respectively.

In order to study the propagation of sound it is
common to take perturbations around a central
value $p_o,\rho_o$
\be\la{pop1}
p(t,x)=p_o(t)+p_1(t,x), \hspace{1cm} \rho(t,x)=\rho_o(t)+\rho_1(t,x)
\ee
where the central values do not depend on the space
coordinates while the perturbations $p_1,\rho_1$
depend on $t$ and $x$.

Keeping first order terms in the pressure and energy
density perturbations one gets by solving the equation
of motion, with a Minkowski metric, the equations
\be\la{pv}
\fr{\pp p_1}{\pp t}=-(\rho+p) \nabla v \hspace{1cm}
\fr{\pp v}{\pp t}= -\fr{1}{\rho+p} \nabla p_1.
\ee
Using the relationship $p=w\rho$
(or equivalently $p_1=w\rho_1$) one has
\be\la{p1r1}
\nabla p_1=w\nabla \rho_1
\ee
which is valid if $w$ is not a function
of $x$. The restriction on $w=p_1/\rho_1=\pp p/\pp \rho$
to not depend on the spatial variations follows
from the hypothesis of adiabatic perturbations.
Eliminating $v$ from  eqs.(\ref{pv}) and using (\ref{p1r1}) we get the
usual  wave equation
\be\la{r1}
\ddot\rho_1-w^2\nabla^2 \rho_1=0.
\ee
Comparing eq.(\ref{r1}) with eq.(\ref{onda}) we
see that the phase velocity is given by $v_p=\vert w\vert=v$
and it is the same for all
  monochromatic wave functions
of the fluid perturbations $\rho_1$
since $b=d=0$. Eq.(\ref{r1}) is valid for a static
(Minkowski) space in the absence of gravitational
interactions.

If we consider a flat FRW space, with metric
$ds^2=dt^2-a(t)^2dx^2$, then the perturbation
equation for radiation with $\delta_r\equiv \delta \rho_r/\rho_r$
is given (in the transverse gauge) by
\be\la{rad}
\ddot\delta_r+\fr{1}{3}k^2 \delta_r=\fr{4}{3} \ddot \delta_c
\ee
where $\delta_c\equiv \delta \rho_c/\rho_c$ is the perturbation
of (cold)  matter and the r.h.s. term in eq.(\ref{rad}) is a source
term for $\delta_r$.  Comparing  eq.(\ref{rad}) and eq.(\ref{onda})
we see that the phase velocity is given by $v^2=1/3$ and in the
absence of the source term, the group  and the phase
velocity would coincide. Since there
is no dissipative term ($d=0$)  the sound velocity
for radiation would be given be $v_p=v_g=v_p=1/\sqrt{3}$.

Notice that even in an expanding universe the perfect
fluid (in this case for radiation) has no dispersion.

\subsection{Scalar Field}

A scalar field is defined by the action
\be\la{lag}
S=\int dx^4\sqrt{-g}\le[\fr{\lm}{2}\le(\partial_\nu\phi \right)^2 - V(\phi)\ri]
\ee
where $g\equiv det[g_{\mu\nu}]$ and we have introduced the parameter $\lm=\pm 1$
in order to incorporate phantom fields in the analysis ($+1$ for a canonical scalar fields and $-1$ for
phantom fields). From
this eq. we can extract the energy momentum tensor
$T_{\mu\nu}(\phi)=\lm\pp_\mu\phi\pp_\nu\phi-g_{\mu\nu}
\le[\half\lm g^{\al\beta}\pp_\al\phi\pp_\beta\phi-V(\phi)\ri]
$ and the equation of motion is
\be\la{mov}
\lm(\pp^\mu\pp_\mu\phi+ 3H\dot\phi)+V'=0.
\ee
where we have set
$[\pp_\mu( g^{\mu\nu}\sqrt{-g})/\sqrt{-g}]\pp_\mu\phi=3H\dot\phi$,
valid in a flat FRW metric,
with $\sqrt{-g}=a(t)^3$, $H=\dot a/a$ the Hubble parameter and  $a(t)$
the scale factor.
The $T^o_o=\rho$ component defines the energy density while
$T^i_i=-p_\phi$ the pressure, for simplicity and
presentation purposes we will only consider perturbations
in one direction $i=x$, giving
\be\la{rp}
  \rp =\lm(\half\dot\phi^2 -\half \pp^i\phi\pp_i\phi )+ V(\phi)
    \hspace{1cm}
  p_\phi =\lm(\half\dot\phi^2 -\half \pp^i\phi\pp_i\phi) -V(\phi)
  \ee
and the equation of state parameter is
\be\la{wpT}
\wp\equiv\fr{p_\phi}{\rp}=\fr{\lm(\half\dot\phi^2 -\half \pp^i\phi\pp_i\phi) - V(\phi)}
{\lm(\half\dot\phi^2 -\half \pp^i\phi\pp_i\phi) + V(\phi)}.
\ee
In cosmology it is usual to assume  the scalar field
to be homogenous in space and therefore $\phi(t,x)=\phi_o(t)$
is only a function of time and the space derivatives
vanish, $\pp_i \phi_o=0$. In such a case eq.(\ref{wpT}) reduces
to eq.(\ref{wp}).

However, contrary to a perfect fluid, in the case of
a scalar field the equation of state cannot be imposed
and it is dynamically determined by the evolution
of $\phi$ given by eq.(\ref{mov}). It is easy to see
that $w$ in eq.(\ref{wpT}) does depend on the
spatial coordinates. Taking a perturbations in the $x$-direction
around the average solution $\phi_o$ one has
\be\la{phoph1}
\phi(t,x)=\phi_o(t)+\varphi(t,x)
\ee
and to first  order in $\varphi$ we have
$V(\phi(t,x))\simeq V_o(\phi_o(t))+V_o'\varphi+O(\varphi^2)$ with $V_o'=\pp V/\pp \phi|_{\phi_o}$
and
$(\dot\phi^2 -\pp^i\phi\pp_i\phi)/2= \dot\phi_o^2/2+\dot\phi_o\dot\varphi+O(\varphi^2)$.
Eq.(\ref{wpT})  up to first order in $\varphi$ gives
\be\la{wptx}
\wp= \wpo+\fr{2\lm\le[V'_o\dot\phi^2_o\varphi(t,x)-2V_o\dot\phi_o\dot\varphi(t,x)\ri]}{p_{\phi o}^2}+O(\varphi^2)
\ee
where $\wpo$ is given by eq.(\ref{wp}) and $p_{\phi o}=\lm\dot\phi_o^2/2-V_o(\phi_o)$.
 The perturbed equation  (\ref{mov})
  up to first order
in $\varphi$ is
\be\la{dvar}
\lm(\ddot\varphi
-\nabla'^2\varphi+3H\dot\varphi)+V(\phi_o)''\varphi=-\fr{1}{2}\dot\phi_o\dot h
\ee
where the r.h.s in eq.(\ref{dvar}) is due to the variation
of the metric tensor $g_{\mu\nu}=\eta_{\mu\nu}+h_{\mu\nu}$
and $\nabla'^2\equiv a^{-2}\nabla^2$ and $V''=\pp^2 V/\pp\phi^2$.
In terms of a Fourier transformation
$\varphi(t,x)=(2\pi)^{-3}\int dk^3 \varphi_k e^{ik\cdot x'/a}$ we can
 write eq.(\ref{dvar})
as
\be\la{dvar1}
\ddot\varphi_k +3H\dot\varphi_k +(\fr{k^2}{a^2}+
\fr{V(\phi_o(t))''}{\lm})\varphi_k=-\fr{1}{2\lm}\dot\phi_{ko}\dot h
\ee
with $k^2=k\cdot k$. Eq.(\ref{dvar1}) gives the wave equation
for the scalar field perturbations and we can see that
the medium is dissipative ($H\neq 0$) and dispersive
($( k^2/a^2+V''/\lm)/k^2$ is $k$ dependent).
In conformal time $a d\tau=dt$ eq.(\ref{dvar1})
becomes
\be\la{dvar2}
\varphi_k^{\star\star} +2 \textbf{H} \varphi_k^\star +(k^2+
\fr{a^2 V(\phi_o(t))''}{\lm})\varphi_k=-\fr{1}{2\lm}\phi_{ko}^\star h^\star
\ee
where $\star$  means derivative w.r.t. $d\tau$ and $\textbf{H}\equiv a^\star/a$.  Comparing eq.(\ref{dvar2})
with eq.(\ref{onda}) we see that $v_p=1$, i.e. the "phase
velocity" is the speed of light, $b=a^2V(\phi_o(t))''/\lm$
and  $d=2\textbf{H}$ are  non zero and time dependent.
Therefore, the wave
function will have a non constant amplitude and the velocity
of each monochromatic wave will differ thus giving
a  group velocity different than the phase velocity.
We also remark that the left hand side of eq.(\ref{dvar2})
is the same for a canonical scalar field with negative
mass square (i.e. a tachyon field with  $V''/\lm=V''<0$) to a phantom field
with positive $m^2$  (i.e. $V''/\lm=-V''<0$). On the other hand, the gravitational
contribution given by the r.h.s of eq.(\ref{dvar2}) has opposite  sign for a canonical field and
a phantom field since $\lm$ changes sign.

\subsection{K-essence}

Let us now consider the following (k-essence)  action for a scalar field \ci{kessence}
\be\la{lkin}
S=\int dx^4\sqrt{-g}K(\phi)F(X)
\ee
with $K(\phi)$ an arbitrary function of the scalar field $\phi$  and $F(X)$
a function of the kinetic term  $X=\pu \phi\pU\phi/2$.  The equation of motion of eq.(\ref{lkin}),
 in a flat FRW metric, is given by
\be\la{eqkin}
\pu \le(a(t)^3F'(X)K(\phi)\pU\phi \ri)-a(t)^3F(X)K'(\phi)=0
\ee
where we have taken $\sqrt{-g}=a^3(t)$ and a prime denotes derivative w.r.t. the argument
and we used $\delta X/\delta\pU\phi=\pu\phi$. Eq.(\ref{eqkin}) is then
\be\la{eqkin2}
\pu\pU\phi+\fr{K'}{K}\pu\phi\pU \phi +\fr{F''}{F'}\pu X\pU \phi+3H\dot\phi-\fr{F}{F'}\fr{K'}{K}=0
\ee
The Lagrangian in eq.(\ref{lkin})
can be reduced to a canonical kinetic term scalar Lagrangian for   $F=X, K=1$ and
eq.(\ref{eqkin2}) reduces to $\pu\pU\phi+3H\dot\phi=0$, i.e. a scalar field without potential.
For an homogenous scalar field $\phi=\phi_o(t)$ we have $\pu X=\dot\phi_o\ddot\phi_o$ and eq.(\ref{eqkin2}) becomes
\be\la{eqkin3}
\pu\pU\phi_o+3H\dot\phi_o \fr{F'}{F'+F''\dot\phi_o^2}+\fr{F}{F'}\fr{K'}{K}\le(\fr{\dot\phi_o^2F-F'}{F'+F''\dot\phi_o^2}\ri)=0
\ee

In order to analyze the evolution of the perturbations we will expand eq.(\ref{eqkin2}) to
first order in $\phi(t,x)=\phi_o(t)+\varphi(t,x)$. Let us defined the functions
$D(\phi)\equiv K'(\phi)/K(\phi),\; E(X)\equiv F''(X)/F'(X)$ and $S(X)\equiv F(X)/F'(X)$
which appear in eq.(\ref{eqkin2}). We expand   the functions $D,E,S$  to first order in $\varphi(t,x)$,
i.e.  $D=D_o+D'\varphi,\;
 E=E_o+E'\delta X,\;S=S_o+S'\delta X$ with $\delta X=\dot\phi_o\dot\varphi$
where the prime denotes derivative w.r.t. the argument.
The  expansion of $X$ and $\pu X$ are given by
\bea\la{x}
X&=&X_o+\delta X=X_o+ \dot\phi_o\dot\varphi+O(\varphi^2)\non\\
\pu X&=&\pu X_o + \dot\varphi\pu\dot\phi_o+\dot\phi_o\pu\dot\varphi +O(\varphi^2)
\eea
with $X_o=\dot\phi^2_o/2,\;\dot X_o=\dot\phi_o\ddot\phi_o$,
 $ \pp_\al X_o=\pp_\al(\pu \phi_o\pU\phi_o/2)=\dot\phi_o\ddot\phi_o\delta_{o\al}$
and  $\pu X\pU \phi=\dot\phi_o \dot X_o
+2\dot\phi_o\ddot\phi_o\dot\varphi+\dot\phi^2_o\ddot\varphi$. To
first order in $\varphi(t,x)$ eq.(\ref{eqkin2}) becomes
\be\la{vkin}
\ddot\varphi-B\nabla'^2\varphi
+3BH\dot\varphi+A\dot\varphi+m^2_{eff}\varphi=-\fr{1}{2}\dot\phi_o\dot
h
 \ee
 where
 \be\la{AB}
 B=1/(1+E_o\dot\phi_o^2)=F'/(F'+F''\dot\phi^2),\hspace{1cm}
A=B(2D_o\dot\phi_o+E'\dot\phi_o^3\ddot\phi_o-S'D_o\dot\phi_o+2E_o\ddot\phi_o
\dot\phi_o)
\ee
 and the effective mass is given by
 \be\la{meff}
m_{eff}^2=B(\dot\phi_o^2-S_o)D'=\fr{(\dot\phi_o^2-S_o)D'}{1+E_o\dot\phi_o^2}.
\ee
Making a fourier transformation of eq.(\ref{vkin}), as in eq.(\ref{dvar1}), we get
\be\la{var-kin} \ddot\varphi_k
+(3BH+A)\dot\varphi_k+(\fr{k^2}{a^2}B+m^2_{eff})\varphi_k=-\fr{1}{2}\dot\phi_o\dot
h
\ee The r.h.s. in eq.(\ref{vkin}) is due to the variation of the
metric as in eq.(\ref{dvar}) and the factor of $B$ in
eqs.(\ref{vkin}) and (\ref{meff}) arises because $\pu X\pU \phi$
involves a second order time derivative of $\varphi$. Comparing
equation (\ref{var-kin})  to eq.(\ref{dvar1}) we see  that they
are similar,
 eq.(\ref{var-kin}) has a second  time derivative term,
a friction term and an effective mass $m^2_{eff}$.
 The friction term,
the effective mass and the phase velocity   $v_p^2=B$, as seen from
eq.(\ref{onda}),
are model and time dependent. The
phase velocity    will, in general, be different than
the  speed of light contrary to a canonical scalar field.
The effective mass square can in principle be positive or negative
and the effect of the friction term is to  damp or enhance the amplitude
of the fluctuations, depending on its sign.

Let us now consider  the particular case presented in  \ci{kessence}.
 In this   special case
$K(\phi)=1/\phi^2$ and $F(X)=g(y)/y$ with $y$ defined as $y=1/\sqrt{X}$ and
 the pressure,  energy density, equation of state and group velocity
are given by \ci{kessence}, \ci{k-MCLT}
\bea
p=\fr{g(y)}{\phi^2 y} &\hspace{1cm}&
\rho=-\fr{g'(y)}{\phi^2}\\
w=-\fr{g(y)}{yg'(y)} &\hspace{1cm}&
v_g =\fr{g(y)-yg'(y)}{y^2g''(y)}.
\la{kvg}\eea
As seen from eqs.(\ref{kvg}) the equation of state parameter (i.e. the phase velocity) diverges
 when $g'(y)=0$ while the
group velocity  diverges at $g''(y)=0$.
In this class of models  we have
$D(\phi)= K'(\phi)/K(\phi)=-2/\phi,\;D'=dD/d\phi=2/\phi_o^2,\;\;
E(X)= F''(X)/F'(X)=y^2(y(g'+yg'')-g)/(2(g-yg'))$
and $S(X)=F(X)/F'(X)=2g/(y^2(g-yg'))$ and
the effective mass of eq.(\ref{meff}) is given by
\be
m^2_{eff}=\fr{4}{y^2\phi_o^2}\fr{[\dot\phi_o^2y^2(g-yg')-2g]}{[2(g-yg')+(y(g+yg')-g)
y^2\dot\phi_o^2]}.
\ee
The coefficients $A, B$ of eq.(\ref{AB}) can  be easily obtained. However,  even in this particular case
the effective mass, $A$ and $B$  have not a simple form and are still model and time dependent but the main
conclusions remain valid.
The sound speed for k-essence differs from a canonical scalar field, its phase velocity is
no longer equal to the speed of light but  and its sound  velocity
is smaller than the speed of light.

\subsection{Curvature Correction}

In the case where we consider corrections to the Einstein
general relativity we have a lagrangian \cite{Capozzolo03,CDTT04}
\be\la{Lcur}
S=\int dx^4\sqrt{-g}\le(f(R) + L_m\ri)
\ee
with $f(R)$ an arbitrary function of 
the Ricci scalar $R$ and $L_m(\phi)$ the matter lagrangian.
The equation of motion of eq.(\ref{Lcur})
 in a flat FRW metric, is given by
\be\la{eqcur}
  f'(R) R_{\alpha\beta} - \frac{1}{2}f(R)g_{\alpha\beta} =
  f'(R)^{;\mu\nu} \left(g_{\alpha\mu}g_{\beta\nu} -
  g_{\alpha\beta}g_{\mu\nu}\right) + \bar{T}_{\alpha\beta}^{M}
\ee
which can be written in a suggestive form \cite{Capozzolo03}
\begin{equation}
   R_{\alpha\beta}  - \frac{1}{2} R g_{\alpha\beta} =
   T^C_{\alpha\beta} + T^M_{\alpha\beta}
\end{equation}
where
\begin{eqnarray}
  T^C_{\alpha\beta} &=& \frac{1}{f'(R)} \left\{\frac{1}{2}\left[f(R) -
  Rf'(R)\right] + f'(R)^{;\mu\nu} \left(g_{\alpha\mu}g_{\beta\nu} -
  g_{\alpha\beta}g_{\mu\nu}\right)\right\}\\
   T^M_{\alpha\beta} &=&  \frac{1}{f'(R)} \bar{T}_{\alpha\beta}^{M}
\end{eqnarray}

In some interesting cases, such as
\begin{equation}
  \begin{array}{cc}
    f(R) = R^{\frac{3}{2}},\qquad &\qquad f(R) = R - \frac{\mu^2}{R}
  \end{array}
\end{equation}
these fourth order equations can be
transformed to the Einstein frame, where the gravitational equations
take the Einstein form, coupled to a scalar field. This is done
through a conformal mapping \cite{Capozzolo03,CDTT04}
\begin{equation}
  \begin{array}{cc}
    \tilde{g}_{\alpha\beta} = F(R) g_{\alpha\beta},\qquad & \qquad\phi
    = G(R)
  \end{array}
\end{equation}
with suitable chosen functions $F(R)$ and $G(R)$.
The resulting equations are
\begin{eqnarray}
  \tilde{R}_{\alpha\beta}  - \frac{1}{2} \tilde{R}
  \tilde{g}_{\alpha\beta} &=&  T^\phi_{\alpha\beta} +
  T^M_{\alpha\beta}\\
  \nabla_\mu\nabla^\mu\phi - V'(\phi) &=& \sigma
\end{eqnarray}
for a suitable source term $\sigma$ \cite{Capozzolo03,CDTT04}. Thus,
in many interesting cases, curvature corrections model of dark energy
can be mapped on the scalar field quintessence paradigm. Therefore,
the small disturbances equations will have the form  (\ref{dvar1}).

It should be noted that if the action (\ref{Lcur}) is treated with a
Palatini variation \cite{cur-Vuc,cur2} the resulting field equations
are of the second order type. The resulting field equations have an
effective cosmological constant which, for a suitable choice of
parameters, can be fitted to reproduce the present deceleration
parameter. Since these equations have the Einstein form, small
perturbations will behave much the same way as in a cosmological model
with a varying cosmological constant. In particular, adiabatic scalar
modes will obey a scalar ave equation, formally similar to (\ref{dvar1}).
However, for lagrangians singular in $R=0$ the newtonian
limit cannot be recovered \cite{Barraco04}.

\section{Solution}\la{sol}

We shall study now the propagation speeds of a scalar field. For
the sake of simplicity, let us consider Eq. (\ref{mov}) in
Minkowski space-time, with $\sqrt{-g} = 1$ in a 1+1 dimensional space-time.
Let us consider the Lagrangian density $
  {\cal L} = \frac{\lambda}{2}[(\pp_{ct} \phi)^2-(\pp_{z} \phi)^2]
   - V(\phi)$, where the constant $\lm$ allow us to include phantom
   fields in the following discussion.
The equation of motion as given by eq.(\ref{mov}) and
the perturbed equation of motion for $\varphi(z,t)$,  eq.(\ref{dvar})
($\phi(z,t)=\phi_0(t)+\varphi(z,t)$), in the absence of the source term,
is
\begin{equation}
  \lambda\left(\frac{1}{c^2}\frac{\partial^2 \varphi}{\partial t^2} -
  \frac{\partial^2 \varphi}{\partial z^2}\right) + m^2 \varphi = 0.
  \label{DiffEq}
\end{equation}
where we have taken $V(\phi_o)''=m^2$. In terms of the Fourier
transformation we have
\be
 \frac{\lm}{c^2}\ddot\varphi_k  + \le(\lm k^2  + m^2\ri) \varphi_k = 0.
  \label{DiffEq2}
\ee
We have not incorporated a friction term because it only enhances or
damps the amplitude of the fluctuations but does not interfere with
the sound speed. Eqs.(\ref{DiffEq}) or (\ref{DiffEq2}) encompasses
several particular cases commonly found in the cosmological
literature: a standard scalar field $ \lambda > 0,\; m^2 > 0 $ with
$1>\wpo>-1$, a negative potential field $ \lambda > 0,\; m^2 < 0$
and a phantom field $ \lambda < 0,\; m^2 > 0$. The last two examples
have regions with $|\wpo|>1$ and/or $|dp_o/d\rho_o|>1$. The second
case corresponds to $\phi$ near a maximum of the potential, which
means the existence of an effective tachyon. It would seem that the
sound velocity becomes grater than $c$ in this case, but this is not
so: the fluid model breaks down when the propagation of perturbations
in the system is considered, even in the ``normal'' case $\mid w \mid
< 1$. The same thing happens for the third case, i.e. phantom field.

We are interested in finding the general solution with the initial
conditions
\be
  \varphi(z,0) =f(z), \hspace{1cm}
  \left(\frac{\partial \varphi}{\partial t}\right)_{t=0} = g(z).
  \label{InitCon}
\ee
It is enough to find the fundamental solution of eq.(\ref{DiffEq})
$\Delta(z,t)$, satisfying
\be
  \Delta(z,0) =0, \hspace{1cm}
  \left(\frac{\partial\Delta}{\partial t}\right)_{t=0} = \delta(z)
  \label{IniDelta}
\ee
since for any set of initial conditions (\ref{InitCon}) we have
\begin{equation}
  \varphi(z,t) = \int^\infty_{-\infty} {d\zeta \left[\dot{\Delta}(z -
  \zeta,t) f(\zeta) + \Delta(z - \zeta,t) g(\zeta)\right]}.
\label{varphi}
\end{equation}
Indeed
\begin{eqnarray}
  \varphi(z,0) &=& f(z)\\
  \lim_{t \to 0} \frac{\partial \varphi(z,t)}{\partial t} &=& \lim_{t \to
  0}  \int^\infty_{-\infty} {d\zeta \left[\ddot{\Delta}(z -
  \zeta,t) f(\zeta) + \dot{\Delta}(z - \zeta,t) g(\zeta)\right]}\\
  &=& \lim_{t \to   0}  \int^\infty_{-\infty} {d\zeta
    \left[\left(\frac{\partial^2\Delta}{\partial z^2} -
  m^2\Delta\right) f(\zeta) +  \delta(z-\zeta) g(\zeta)\right]}\\
  &=& g(z).
\end{eqnarray}
Using a Fourier decomposition for $\Delta$
\begin{equation}
  \Delta(z,t) = \frac{1}{\sqrt{2\pi}} \int^\infty_{-\infty} {dk
  e^{ikz} \Delta(k) e^{i\omega t}}
\end{equation}
we find the dispersion relation
\begin{equation}
   \omega = \pm c \sqrt{k^2 + \mu^2} \label{RelDisp}
\end{equation}
with $\mu^2\equiv m^2/\lambda$. Observe that negative kinetic
energy, $\lambda<0$,  behaves as a negative $m^2$ with respect
to propagation. Using the initial conditions eqs.(\ref{IniDelta}) we
find
\begin{equation}
   \Delta(z,t) = \frac{1}{2\pi} \int^\infty_{-\infty} {dk
  \frac{\sin\omega(k)t}{\omega(k)} e^{ikz}  }.
\end{equation}
Now we use the integral representation \cite[(Integral 6.677.6)]{GR4e}
\begin{eqnarray}
  \frac{\sin ct\sqrt{k^2 + \mu^2}}{ c\sqrt{k^2 + \mu^2}} &=&
  \frac{1}{2c} \int^{ct}_{-ct} d\xi J_0\left(\mu\sqrt{(ct)^2 -
  \xi^2}\right) e^{-ik\xi}\la{sinct1} \\
   &=& \frac{1}{2c} \int^{\infty}_{-\infty} d\xi
  J_0\left(\mu\sqrt{(ct)^2 -   \xi^2}\right)\Theta[(ct)^2 -
  \xi^2]  e^{-ik\xi}
  \label{sinct}
\end{eqnarray}
where  $\Theta[x]$ is the Heaveside function ($\Theta[x]=1$ for $x
\geq 0$ and $\Theta[x]=0, x <0$) and $J_0(x)$ the Bessel function
of zero order (notice that the change in integration from eq.(\ref{sinct1})
to eq.(\ref{sinct}) is taken care by the Heaveside function), to
obtain
\begin{eqnarray}
  \Delta(z,t) &=& \frac{1}{4\pi c} \int^\infty_{-\infty} dk e^{ikz}
  \int^{ct}_{-ct} J_0\left(\mu\sqrt{(ct)^2 -
  \xi^2}\right) e^{-ik\xi} d\xi\la{delta1}\\
  &=& \frac{1}{4\pi c} \int^{\infty}_{-\infty} dk d\xi
  J_0\left(\mu\sqrt{(ct)^2 -
  \xi^2}\right) \Theta[(ct)^2 -
  \xi^2]  e^{ik(z-\xi)}\la{delta2}\\
  &=& \frac{1}{2c}  J_0\left(\mu\sqrt{(ct)^2 - z^2}\right)
  \Theta[(ct)^2 - z^2] \label{eq:Delta}.
\end{eqnarray}
The last equation (\ref{eq:Delta}) is valid for  $t\neq 0$ and it
gives $\Delta(z,t=0)=0$, as can be seen from   eq.(\ref{delta2}).
Taking the derivative of eq.(\ref{eq:Delta}) we
have
\bea
\dot\Delta(z,t)= &-&\frac{\mu c t }{2} \frac{J_1\left(\mu\sqrt{(ct)^2 -
  z^2}\right)}{\sqrt{(ct)^2 -
  z^2}} \Theta[(ct)^2 -z^2]\non \\
   &+& \frac{1}{2} J_0\left(\mu\sqrt{(ct)^2 -
  z^2}\right)\left(\delta(ct-z)+\delta(ct+z)\right)
  \label{dotD}
\eea
and for $t=0$ we get $\dot\Delta(z,0)=\delta(z)$.
 Eqs. (\ref{eq:Delta}) and (\ref{dotD})  are the main results of this
section. They show that a signal propagates always with speed
smaller or equal than $c$, since $\Delta(z,t)$ and
$\dot\Delta(z,t)$ vanish for $|z| > ct$. This ensures that a
signal generated at $t=0, z_0$ will necessarily arrive at a point
$|z-z_0|$ at a time larger or equal than $ct$, as can be seen from
the vanishing of the wave function $\varphi(z,t)$ for $|z-z_0|>
ct$ from eqs.(\ref{varphi}), (\ref{eq:Delta}) and (\ref{dotD}).

On the other hand, the result for $m^2 < 0$ (or $\lambda < 0$)
can be obtained easily from Eq.  (\ref{sinct}) or (\ref{eq:Delta})
using analytic continuation in the parameter $\mu = m/\sqrt{\lambda}$. In
both cases one gets
\begin{equation}
  \Delta_I(z,t) =  \frac{1}{2c}  I_0\left(|\mu|\sqrt{(ct)^2
  - z^2}\right)
  \Theta[(ct)^2 - z^2] \label{eq:Delta_I}
\end{equation}
where  $I_0$ is the modified Bessel function of zero order.  The
amplitude of eq.(\ref{eq:Delta_I}) grows exponentially with time,
showing explicitly the instability of a field with  $m^2 < 0$ (or
$\lambda < 0$). However, it also shows that the speed of the
wave packet is smaller or equal to the speed of light, as for the
standard scalar field case ($m^2>0,\lambda>0$). The instability
growth of the perturbation is ameliorated by the expansion of the universe
since instead of  growing exponentially fast it grows power like.
The increase of the amplitude implies that the first order approximation
in the perturbed differential equation (\ref{DiffEq}) will cease to be valid.
This implies that we can no longer use eq.(\ref{DiffEq}) to study the
evolution of $\varphi$ and non-perturbative methods must be used.
In this case the universe will be become inhomogeneous. However,
we would like to emphasis that the sound speed never becomes larger
than the speed of light.

It should be stressed that the above results are general. Indeed, the
Principle of Equivalence guarantees that in a freely falling reference
systems, the equations of motion for the fluid or scalar field will
take their Minkowskian form.

\begin{figure}[tbp]
  \begin{center}
    \includegraphics{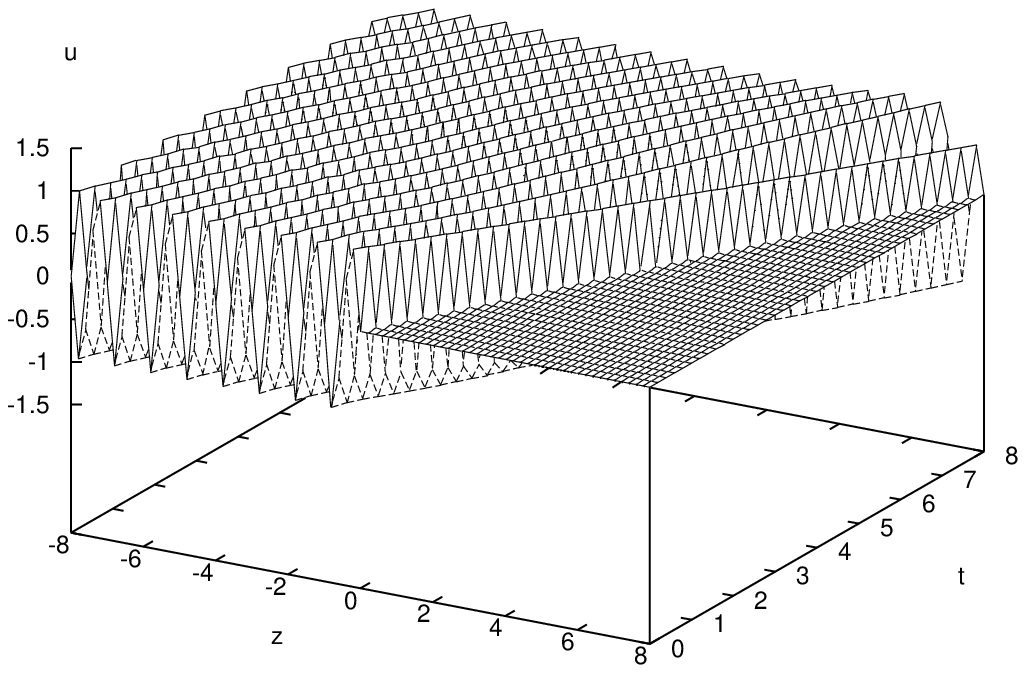}\\
    \includegraphics{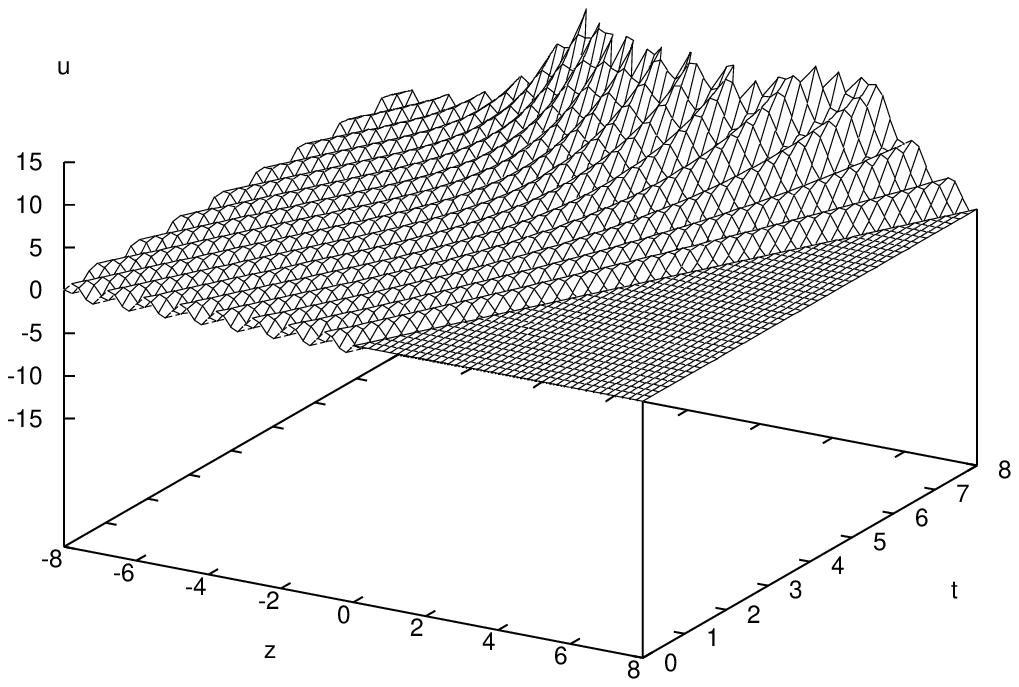}
  \end{center}
  \caption{Propagation of a semi-infinite wave train of frequency
  $\omega=2\mu$, moving to the right.  Fig.a shows the amplitude of
  the pulse as a function of $(z,t)$ in the normal  $\mu^2 >
  0$ case.  Fig.b displays the propagation
  of the same initial train in the anomalous (tachyonic) case $\mu^2 <
  0$. In this latter case, the signal shows exponential growth but in both
  cases the amplitude  vanishes outside the lightcone, i.e. for $z>t$.}
  \label{fig:mu=2}
\end{figure}

This behavior can be simply illustrated solving eq.(\ref{DiffEq}) in
1+1 dimensional space-time. We assume that a rectangular semi-infinite
wave train of unit amplitude is incident at the spatial point $z = 0$ and
look its subsequent behavior. The
initial conditions for such a pulse are
\begin{eqnarray*}
  f(z) &=& sin(kx)*\Theta(-x)\\
  g(z) &=& - \frac{d\,f(z)}{d\,z}
\end{eqnarray*}

 Fig.2a shows the normal case with $\mu^2 > 0$ and $E = \omega =2
\mu$. The front of the train moves towards the right with a speed
equal to $c$, but the first crest diminishes and settles
slowly into the group velocity. This process is just beginning in
the figure. The field is stable and there is no increase of the
amplitude with time.

Fig.2b shows the propagation of the same wave train in the case
$\mu^2 < 0$, i.e. when the field is ``tachyonic''. In spite of
that character, the signal velocity is equal to $c$ and the
amplitude of the wave vanishes outside the lightcone (i.e. for
$z>t$).  However, the system is
 unstable and shows exponential growth of the signal as can be
seen from the height of the amplitude around $z=0, t=8$.

\section{Conclusions}
\label{Conclusions}

We have shown that scalar fields have naturally regions where
$|\wpo|>1$ and/or $|dp_o/d\rho_o|>1$ and these properties seem to
contradict causality. In some of these regions the scalar field can be
interpreted as an effective tachyon (i.e. $m^2<0$) or phantom
(i.e. $\lm\dot\phi^2<0$) scalar field, i.e. $\mu^2<0$ in both cases. We studied the evolution
of the scalar field perturbations for models with
$|\wpo|>1$ and/or $|dp_o/d\rho_o|>1$ including tachyon, phantom,
k-essence and curvature models. We have shown that
even in these extreme cases   the scalar field is consistent with causality
and Lorentz invariance provided dispersion is taken into account
and the sound speed is never larger than the speed of light.
Dispersion, however, appears naturally for scalar fields having
nontrivial potentials (e.g. mass term) such as those used to model
dark energy. However, when $\mu^2$ becomes negative the amplitude of the perturbations
grows rapidly and the first order approximation breaks down.
The universe becomes, therefore, inhomogeneous and non-perturbative
mathematical methods must be used to study the evolution
of the perturbations.

\begin{acknowledgments}
This work was supported in part by CONACYT project 32415-E and
DGAPA, UNAM project IN-110200.
\end{acknowledgments}


\begin{thebibliography}{99}

\bib{CMB} {P. de Bernardis {\it et al}. Nature, (London) 404:955
  (2000) S. Hannany {\it et al}.,Astrophys.J.545 L1-L4  (2000)}

\bib{SN1a} A.G. Riess {\it et al.}, Astron. J. 116:1009  (1998); S.
Perlmutter {\it et al}, ApJ 517:565  (1999); P.M. Garnavich {\it et
al}, Ap.J 509:74 (1998); J.L. Shievers at al,Astrophys.J. 591:590 (2003)

\bib{sn} S. Hannestad , E. Mortsell,    Phys.Rev. D66:063508  (2002)

\bib{neww} Carlo Baccigalupi, Amedeo Balbi, Sabino Matarrese,
Francesca Perrotta, Nicola Vittorio,  Phys.Rev. D65:063520  (2002)

\bib{insta} P.H. Frampton hep-th/0302007

\bib{carter} B. Carter gr-qc/0205010

\bib{cald} R.R. Caldwell, Phys.Lett. B545:23 (2002).

\bib{kessence} C. Armendariz-Picon, V. Mukhanov and P.J. Steinhardt,
Phys.Rev.Lett.85,4438 (2000); Phys.Rev.D63,103510 (2001)

\bib{k-MCLT} M. Malquarti, E. Copeland, A. Liddle, M. Trodden,
 Phys.Rev.D67:123503,2003

\bib{cur-Vuc} D. Barraco, V.H. Hamity and H. Vucetich,
Gen. Rel. Grav. 34 (2002) 533

\bib{cur1} S. Capozziello, V.F. Cardone, S. Carloni, A. Troisi.
 Int. J. Mod. Phys. D12:1969-1982,2003

\bibitem{cur2} D. Vollick, Phys.Rev.D68 (2003) 063510

\bibitem{Capozzolo03}  S. Capozziello,  S. Carloni, A. Troisi, "Recent
  Research Developments in Astronomy \& Astrophysics"-RSP/AA/21-2003

\bibitem{CDTT04}  S. M. Carroll, V. Duvvuri, M. Trodden, M. S. Turner,
  astro-ph/0306438

\bibitem{Barraco04}  A. E. Dominguez, D. E. Barraco, Phys. Rev. D70
  (2004) 043505

\bib{mcin}B. McInnes,  astro-ph/0210321; JHEP 0208:029 (2002)

\bib{cyc} P.J. Steinhardt, N. Turok, Phys.Rev. D65:126003 (2002)

\bib{ekp} J. Khoury,   B. A. Ovrut, P. J. Steinhardt, N. Turok,
 Phys.Rev. D64:123522 (2001)

\bib{linde} A. Linde, JHEP 0111:052,2001; N. Felder, A.V. Frolov, L.
Kofman, A. V. Linde,  Phys.Rev. D66:023507 (2002)

\bib{ax.neg} A. de la Macorra, G. German astro-ph/0212148

\bibitem{Schweber} S. S. Schweber, \emph{An introduction to the
  theory of quantum fields} (Harper and Row, New York, 1960).

\bibitem{Stratton} J. A. Stratton, \emph{Electromagnetic Theory}
  (McGraw-Hill, New York, 1941).

\bibitem{sinct}{c.f. G-R:3.876}
\bibitem{GR4e} I. S. Gradshteyn and I. M. Ryzhik, \emph{Tables of
  Integrals, Series and Products} 4th ed. (Academic Press, New York,
  1965)

\bibitem{Sommerfeld} A. Sommerfeld, Ann. Physik {44}:177 (1914)

\bibitem{Brillouin}  L. Brillouin, Ann. Physik {44}:203 (1914)



\end{thebibliography}
\end{document}